\documentclass{article}

\usepackage{color}
\usepackage[latin1]{inputenc} 
\usepackage[english]{babel} 
\usepackage[T1]{fontenc}
\usepackage{ae}
\usepackage{amsmath, amssymb, amsfonts}
\usepackage{mathrsfs}
\usepackage[mathcal]{euscript}
\usepackage{graphicx, psfrag}
\usepackage{subfigure}
\usepackage[official]{eurosym}

\title{
		A simple proof of the Jarzynski equality?
		}
\author{
		F.~Douarche\\
		{\it
		Laboratoire de Physique de l'ENS Lyon --- CNRS UMR 5672
		}\\
		{\it
		46, All\'ee d'Italie --- 69364 Lyon Cedex 07, France
		}
		}
		
\def\De{\Delta}

\def\lam{\lambda}
\def\lam{\lambda}

\def\mr{\rangle} 
\def\Ml{\left\langle} 

\def\Mr{\right\rangle} 

\def\d{\mathrm{d}} 

\def\D{\partial}

\def\e{\mathrm{e}}

\begin{document}

\maketitle


\section{Introduction}

In this short communication, I give a very simple derivation of the Jarzynski
equality \cite{JE}
\begin{equation}
	\De F = -\beta^{-1} \ln{\Ml \e^{-\beta W} \Mr},
	\label{JE}
\end{equation}
which allows to compute the free energy difference $\De F = F_B - F_A$ of a
body, which is driven between two equilibrium states $A$ and $B$ by an external
(time-dependent) force $\lam$, from the probability distribution function
${\rm P}(W)$ of the work done on the system
\begin{equation}
	W = \int_{t_A}^{t_B} \frac{\D H_t(x, \lam)}{\D t}\, \d t
	= \int_{t_A}^{t_B} \frac{\D H_t(x, \lam)}{\D \lam}\, \dot{\lam}\, \d t,
	\label{WORK}
\end{equation}
regardless of the nature of the transformation (reversible or irreversible)
between the states $A$ and $B$.
Here $H_t(x, \lam)$ denotes the full (time-dependent) Hamiltonian of the system,
which is supposed to be in thermal contact with a heat bath of temperature
$T = 1 / k_B \beta$, the whole ensemble being isolated, $x = (p, q) = (p_1,
\dots, p_n, q^1, \dots, q^n)$ denotes the phase-space point of the system (which
is supposed to have $2n$ degrees of freedom), and $\Ml f(W) \Mr = \int f(W)\,
{\rm P}(W)\, \d W$ denotes an average over the ensemble of realizations of the
work $W$ done on the system (for any ``smooth'' function $f$).

There are two parts in this derivation.
The first is mathematical and will be discussed later.
The result is the following: let $E(x)$ be the (time-independent) energy of a
body in thermal equilibrium with a heat bath of temperature $T$.
If its energy can be written in the additive manner $E(x) = E_0(x) + V(x)$,
which can always be achieved by making certain very general assumptions, then
one can compute the free energy $F$ of the body (and all the relevant
thermodynamic quantities) from the only knowledge of the equilibrium Gibbsian
distribution $\rho_0(x) = Z_0^{-1}\, \e^{-\beta E_0(x)}$, where $Z_0 = \int
\e^{-\beta E_0(x)}\, \d x$ is the partition function corresponding to the
``unperturbed'' energy $E_0(x)$ of the body (and to its ``unperturbed'' free
energy $F_0 = -\beta^{-1}\, \ln{Z_0}$).

The second part of the argument, which relates the perturbation energy $V$ to
the thermodynamical work $W$ done on the system, first in the time-independent
case, and then in the time-dependent case, is physical.


\section{Thermodynamic perturbation theory}

The fundamental need for perturbation theory runs as follows: for example,
in quantum mechanics, the goal is to compute the eigenvalues $\lam$ (the energy
spectrum) and the eigenvectors $u$ (the wave functions) of the (time-dependent
or not) Hamiltonian $H(x)$ of a system, from the eigenvalue equation $H(u) =
\lam\, u$, where $H$ is generally a very complicated differential operator,
occasionally depending on time via some time-dependent energy coupling (due,
e.g., to some external time-dependent force).

However, for evident analytical reasons, the calculus of the eigenvalues $\lam$
and of the eigenvectors $u$ is generally intractable.
In order to circumvent this problem, one generally write (when this is possible)
the energy of the system in the additive manner $H(x) = H_0(x) + V(x)$, where
$H_0(x)$ denotes the ``unperturbed'' energy of the system (i.e. the quantity
from which one can compute the eigenvalues $\lam_0$ and the eigenvectors $u_0$),
and $V(x)$ the perturbation energy.
Then, one computes by iteration $\lam$ and $u$ from the only knowledge of the
``unperturbed'' eigenvalues $\lam_0$ and eigenfunctions $u_0$, which are
computable \cite{QM}.
Eventually, the perturbation energy $V(x)$ can depend on some external force
$\lam$, time-dependent or not (e.g., $\lam$ can be an electrical or magnetic
field).

The situation is very similar in statistical mechanics: once ones has set up the
Hamiltonian $E(x)$ of the system (which, again, is in thermal contact with a
heat bath of temperature $T$, the whole ensemble being isolated), the
fundamental problem is to compute the partition funtion $Z$ of the system
\[
	Z = \int \e^{-\beta E(x)}\, \d x,
\]
where $\d x = \d p\, \d q = \prod \d p_i\, \d q^i$, from which one can compute
the free energy $F$ of the system
\[
	F = -\beta^{-1} \ln{Z},
\]
and, e.g., the thermal average energy of the system $\Ml E \Mr =
-\frac{\D \ln{Z}}{\D \beta}$ \cite{SM}.

Again, these well-known formulas are purely formal, and inapplicable in
realistic cases, because the energy $E(x)$ of the body is generally a very
complicated function of the coordinates $q$ and momenta $p$ of the ``particles''
constituting the body, and the integration over the phase-space cannot be
carried out in order to compute $Z$ [in the quantum case, this calculus is of
combinatoric nature, and requires a counting of the (occasionally degenerate)
states $|n \mr$ of the system via the formula $Z = \sum \e^{-\beta E_n}$, which
also becomes quickly an impossible task].
However, as we will show below, much more can be said about $Z$, $F$ and
$\Ml E \Mr$, at least formally.\\

Let us first assume that the energy of the body (which, again, is in thermal
contact with a heat bath of temperature $T$, the whole ensemble being isolated)
is time-independent, and that it can be written in the additive manner
\begin{equation}
	E(x) = E_0(x) + V(x),
	\label{EQ1}
\end{equation}
where ``everything'' can be calculated from the ``unperturbed'' energy $E_0(x)$
of the body [the partition function $Z_0 = \int \e^{-\beta E_0(x)}\, \d x$, the
free energy $F_0 = -\beta^{-1} \ln{Z_0}$, etc.], but not from its ``full''
energy $E(x)$.
Therefore, the fundamental problem is to devise a method to compute, even
approximatively, the partition function $Z$, the free energy $F$, etc.,
corresponding to the ``full'' energy $E(x)$ of the system.
Indeed, we will show that if Eq.\,(\ref{EQ1}) holds, then one can compute
exactly $Z$ and $F$, without anymore assumptions than Eq.\,(\ref{EQ1}), which
relies on the separability of the Hamiltonian of the system, and on thermal
equilibrium.

From these very few assumptions, the algebra is very simple.
Let us first compute the partition function $Z = \int \e^{-\beta E(x)}\, \d x$
of the system.
From Eq.\,(\ref{EQ1}), one gets
\begin{equation}
	Z = Z_0 \int \rho_0(x)\, \e^{-\beta V(x)}\, \d x
	= Z_0\, \Ml \e^{-\beta V} \Mr_0,
	\label{EQ2}
\end{equation}
where $\rho_0(x) = Z_0^{-1}\, \e^{-\beta E_0(x)}$ is the ``unperturbed''
equilibrium Gibbsian distribution of the system.
Therefore, the free energy of the system is given by
\begin{equation}
	F = -\beta^{-1} \ln{Z}
	= F_0 - \beta^{-1}\, \ln{\Ml \e^{-\beta V} \Mr_0}.
	\label{EQ3}
\end{equation}

This result was derived up to the second order [i.e. only with the first two
cumulants of $V$ in Eq.\,(\ref{EQ3})] by Peierls and Bogoliubov, who gave an
upper bound for the free energy $F$ of the system \cite{SM} [indeed,
Eq.\,(\ref{EQ3}) indicates that the ``full'' free energy $F$ is merely the
cumulants generating function of the perturbation energy $V$].
Our result is much stronger, and looks amazingly similar to the Jarzynski
equality (\ref{JE}).


\section{Dynamical interpretation}

The preceeding result holds for a time-independent and equilibrium system.
However, provided that the system is initially (say, at time $t = 0$) at
equilibrium, these results are also true if the energy of the system depend on
time through a time-dependent perturbation energy, which can be due to an
external force $\lam$ which is switched on at times $t >0$.
Therefore, if the energy of the system reads
\begin{equation}
	E_t(x, \lam) = E_0(x) + V_t(x, \lam),
	\label{EQ4}
\end{equation}
where the time-dependent perturbation energy $V_t$ is only non-zero for times
$t > 0$, then one trivially has
\begin{equation}
	F_t	= F_0 - \beta^{-1}\, \ln{\Ml \e^{-\beta V_t} \Mr_0}
	\quad \textrm{for all} \quad t > 0,
	\label{EQ5}
\end{equation}
which has exactly the same meaning as Eqs.\,(\ref{EQ2})-(\ref{EQ3}).

In fact, the perturbation energy $V_t$ can be identified to the work $W_t$ done
on the system on the time interval $(0, t)$ in a very simple way [one simply has
to put $t_A = 0$ and $t_B = t$ in Eq.\,(\ref{WORK}), but here $B$ does not
necessarily denotes an equilibrium state].
Since the energy $E_t$ of the system depends explicitly on time only through the
external force $\lam(t)$, one has by applying the chain rule
\[
	\frac{\d E_t(x, \lam)}{\d t} = \frac{\D E_t(x, \lam)}{\D t}
	= \frac{\D E(x, \lam)}{\D \lam} \frac{\d \lam}{\d t}.
\]
Therefore, by Eq.\,(\ref{WORK}) one has $W_t = E_t - E_0 = V_t$, and one finally
gets
\begin{equation}
	F_t	= F_0 - \beta^{-1}\, \ln{\Ml \e^{-\beta W_t} \Mr_0},
	\label{EQ6}
\end{equation}
which resembles amazingly the Jarzynski equality (\ref{JE}).


\section{Discussion and conclusion}

The formula (\ref{EQ6}) obtained above is not really the Jarzynski equality
(\ref{JE}), since the Jarzynski equality computes the free energy difference of
the system from the work distribution ${\rm P}(W)$, which takes into account all
the (in principle) nonequilibrium fluctuations of the work $W$ done on the
system during the transformation between the (equilibrium) states $A$ and $B$.
In our derivation, this is obviously not the case, since the average $\Ml \cdot
\Mr$ is performed from the equilibrium Gibbsian distribution $\rho_0(x) =
Z_0^{-1}\, \e^{-\beta E_0(x)}$.

However, this slight detail perhaps throws some light on the debate started by
Cohen and coworkers \cite{COHEN}, and perhaps explains why all the experiments
which has been performed in order to test the Jarzynski equality are successful,
even in very defavorable (irreversible) cases (and more strikingly, when the
state $B$ is not an equilibrium one) \cite{MOI}.

To conclude, we should add that among the hypotheses needed to the result
(\ref{EQ6}), the ergodicity of the system is of course required.
However, even unrealistic, this is a rather weak statement, because we only need
ergodicity at times $t \leq 0$, that is to say without external (time-dependent)
driving force $\lam$.\\


\noindent{\sl Acknowledgments} ---
The author thank S.~Ciliberto and K.~Gaw\c edzki for useful discussions.



\end{document}